  \documentstyle[preprint,aps]{revtex} 
  \tightenlines                                    
\begin{document}
\draft
\preprint{}
\title{Controlling Quantum State Reduction}
\author{Masanao Ozawa}
\address{School of Informatics and Sciences,
Nagoya University, Chikusa-ku, Nagoya 464-01, Japan}
\date{\today}
\maketitle
\begin{abstract}
Every measurement leaves the object in a family of states indexed by
the possible outcomes.
This family, called the posterior states, is usually 
a family of the eigenstates of the measured observable, 
but it 
can be 
an arbitrary family of states by 
controlling the object-apparatus interaction.
A potentially realizable object-apparatus interaction measures
position in such a way that the posterior states are the translations 
of an arbitrary wave function. 
In particular, position can be measured without perturbing the object
in a momentum eigenstate.
\end{abstract}

\pacs{PACS numbers: 03.65.Bz, 03.67.-a, 04.80.Nn}
\narrowtext

From a statistical point of view a quantum measurement is specified
by the {\em outcome distribution}, the probability distribution of 
the outcome, and the {\em state reduction}, the object state change 
from the state just before measurement to the state 
just after measurement conditional upon the outcome.
If two measurements have the same outcome distribution and the same
state reduction, they are {\em statistically equivalent}.
On the other hand, a measuring process consists of two stages; 
the {\em first stage} is the interaction between the object and
the apparatus that transduces the measured observable to the
probe observable and the {\em second stage} is the detection of the
probe observable which is amplified thereby to a 
directly-sensible, macroscopic outcome variable without disturbing the
object system.

The conventional derivation of the state reduction caused by a given 
apparatus is to compute the state of the object-apparatus composite
system just after the first stage due to the Schr\"{o}dinger equation
for the composite system and to apply the projection postulate to the
subsequent probe detection \cite{Kra71}. 

The validity of this derivation is, however, limited or even
questionable, because of the following reasons:

1. The probe detection, such as photon counting, in some measuring
apparatus does not satisfy the projection postulate \cite{IUO90}.

2. When the probe observable has continuous spectrum, the projection
postulate cannot be formulated properly in the standard formulation
of quantum mechanics \cite{84QC}.

3. Another measurement on the same object can follow immediately
after the first stage so that the state reduction should determine
the state just after the first stage, but the projection postulate 
for the probe detection gives the state just after the second stage
\cite{98CQPPT}.

A rigorous and consistent, alternative derivation of the state
reduction without appealing to the
projection postulate has been established in \cite{84QC,85CA}.
Based on this approach, the statistical equivalence classes of all
the possible quantum measurements have been characterized 
as the normalized completely positive map valued measures in \cite{84QC}. 
This result shows quite generally that the state reductions of
measurements of the same observable can occur in surprisingly
rich variety of ways.
In order to exploit such richness for precision measurement 
and quantum information,
in this paper I will investigate the controllability of
state reduction. 

After presenting the general theory of measurement statistics,
it will be shown that for any Borel function $x\mapsto\rho_{x}$ 
from the real line to the density operators
there is a model of measurement of any observable such that 
the measured object is left in the state $\rho_{x}$ just after 
measurement with outcome $x$ regardless of the prior object state.
A potentially realizable interaction will be found also for the model 
of position measurement such that the object is left in 
$\rho_{x}=e^{-ixP/\hbar}|\phi\rangle\langle\phi|e^{ixP/\hbar}$,
where $P$ is the momentum observable
and $\phi$ is an arbitrarily given wave function.

Let ${\bf S}$ be an object to be measured and let ${\bf A}$ be an apparatus
measuring an observable $A$ of ${\bf S}$.
The Hilbert spaces of ${\bf S}$ and ${\bf A}$ are denoted by ${\cal H}$ 
and ${\cal K}$, respectively.
The process of measurement of $A$ at time $t$ using ${\bf A}$
is described as follows.
The interaction between ${\bf S}$ and ${\bf A}$ are turned on 
from the time $t$
to a later time $t+\Delta t$ so that after $t+\Delta t$ the object is 
free from the apparatus.
The outcome of measurement is obtained by detecting the probe observable,
denoted by $B$, in ${\bf A}$ at the time $t+\Delta t$.
This detection process takes time $\tau$ so that at the time 
$t+\Delta t+\tau$ the observer can read out directly the outcome variable.
The time $t$ is called the {\em time of measurement},
the time $t+\Delta t$ is called the {\em time just after measurement},
and the time $t+\Delta t+\tau$ is called the {\em time of readout}.

For any Borel set $\Delta$ in the real line ${\bf R}$, 
the spectral projection of $A$ corresponding 
to $\Delta$ is denoted by $E^{A}(\Delta)$.
Suppose that the time evolution of the composite system ${\bf S}+{\bf A}$
from $t$ to $t+\Delta t$ is represented by a unitary operator $U$ on
the Hilbert space ${\cal H}\otimes{\cal K}$.
Suppose that ${\bf S}$ is in an arbitrary state $\rho(t)$ 
and that ${\bf A}$ is prepared in a fixed state $\sigma$ at the time $t$.
Then, ${\bf S}+{\bf A}$ is in the state $U(\rho(t)\otimes\sigma)U^{\dagger}$
at the time $t+\Delta t$.
Let ${\bf a}(t)$ be the outcome variable of the apparatus;
the time parameter $t$ represents the time of measurement.
Since the outcome is obtained by the probe detection at $t+\Delta t$, 
according to the Born statistical formula
the outcome distribution, i.e., the probability distribution of ${\bf a}(t)$, 
is given by 
\begin{equation}\label{eq:215a}
{\rm Pr}\{{\bf a}(t)\in\Delta\}
={\rm Tr}[(I\otimes E^{B}(\Delta))U(\rho(t)\otimes\sigma)U^{\dagger}]
\end{equation}
for any Borel set $\Delta$.
The apparatus is, therefore, modeled by the quadruple
$({\cal K},\sigma,U,B)$ of a Hilbert space ${\cal K}$, 
a density operator $\sigma$ on ${\cal K}$, 
a unitary operator $U$ on ${\cal K}\otimes{\cal H}$,
and a self-adjoint operator $B$ on ${\cal K}$, where
${\cal K}$ represents the state space of the apparatus,
$\sigma$ the {\em preparation} of the apparatus,
$U$ the {\em interaction} between the object and the apparatus,
and $B$ the {\em probe} to be detected. 

The requirement for the model apparatus $({\cal K},\sigma,U,B)$ to describe 
a measurement of the observable $A$ is that the outcome distribution is 
identical with the probability distribution of $A$ in the state 
$\rho(t)$, i.e., 
\begin{equation}\label{eq:215b}
{\rm Pr}\{{\bf a}(t)\in\Delta\}={\rm Tr}[E^{A}(\Delta)\rho(t)]
\end{equation}
for any Borel set $\Delta$.

For any Borel set $\Delta$, let $\rho(t+\Delta t|{\bf a}(t)\in\Delta)$ 
be the state
at $t+\Delta t$ of ${\bf S}$ conditional upon ${\bf a}(t)\in\Delta$; 
if the object ${\bf S}$ is sampled randomly from the subensemble 
of the similar systems that yield the outcome of the 
$A$-measurement in the Borel set $\Delta$, then ${\bf S}$ is 
in the state $\rho(t+\Delta t|{\bf a}(t)\in\Delta)$ at the time $t+\Delta t$.
Since the condition ${\bf a}(t)\in{\bf R}$ makes no selection, the state
change $\rho(t)\mapsto\rho(t+\Delta t|{\bf a}(t)\in{\bf R})$ is called the
{\em nonselective state change} and when $\Delta\ne{\bf R}$ the state change
$\rho(t)\mapsto\rho(t+\Delta t|{\bf a}(t)\in\Delta)$ is called the 
{\em selective state change}.

According to the standard argument,
the nonselective state change is determined by
\begin{equation}  
\rho(t+\Delta t|{\bf a}(t)\in{\bf R})=
{\rm Tr}_{{\cal K}}[U(\rho(t)\otimes\sigma)U^{\dagger}],
\end{equation}
where ${\rm Tr}_{{\cal K}}$ denotes the partial trace over ${\cal K}$.

In order to determine the selective
state change caused by the apparatus ${\bf A}$, 
suppose that at the time $t+\Delta t$ just after measurement
the observer were to measure an arbitrary observable $X$
of the same object ${\bf S}$.  Let ${\bf X}$ be the apparatus measuring $X$
and let ${\bf x}$ be its outcome variable.
Since the $B$-measurement at the time $t+\Delta t$ does not disturb
the object ${\bf S}$, the joint probability distribution of the outcome
of the $B$-measurement and the outcome of the $X$-measurement
satisfies the joint probability formula for the
simultaneous measurement of $I\otimes B$ and 
$X\otimes I$ in the state 
$U(\rho(t)\otimes\sigma)U^{\dagger}$ \cite{98OSPPT}.
It follows that the joint probability distribution of the outcome
${\bf a}(t)$ of the apparatus ${\bf A}$ and the outcome ${\bf x}(t+\Delta t)$
of the apparatus ${\bf X}$ is given by
\begin{eqnarray}
\lefteqn{{\rm Pr}\{{\bf a}(t)\in\Delta,{\bf x}(t+\Delta t)\in\Delta'\}}
\quad\nonumber\\
&=&{\rm Tr}[(E^{X}(\Delta')\otimes E^{B}(\Delta))
U(\rho(t)\otimes\sigma)U^{\dagger}].
\label{eq:215d}
\end{eqnarray}
On the other hand, using the state $\rho(t+\Delta t|{\bf a}(t)\in\Delta)$
the same joint probability distribution can be represented by
\begin{eqnarray}
\lefteqn{{\rm Pr}\{{\bf a}(t)\in\Delta,{\bf x}(t+\Delta t)\in\Delta'\}}
\quad\nonumber\\
&=&
{\rm Tr}[E^{X}(\Delta')\rho(t+\Delta t|{\bf a}(t)\in\Delta)]
{\rm Pr}\{{\bf a}(t)\in\Delta\}.
\label{eq:215e}
\end{eqnarray}
From (\ref{eq:215a}), (\ref{eq:215d}), and (\ref{eq:215e}), the state 
$\rho(t+\Delta t|{\bf a}(t)\in\Delta)$ is uniquely determined as
\begin{eqnarray}
\lefteqn{\rho(t+\Delta t|{\bf a}(t)\in\Delta)}\quad\nonumber\\
&=&
\frac{{\rm Tr}_{{\cal K}}[(I\otimes E^{B}(\Delta))
U(\rho(t)\otimes\sigma)U^{\dagger}]}
     {{\rm Tr}[(I\otimes E^{B}(\Delta))U(\rho(t)\otimes\sigma)U^{\dagger}]}
\label{eq:215f}
\end{eqnarray}
for every Borel set $\Delta$ with ${\rm Pr}\{{\bf a}(t)\in\Delta\}>0$.
The above formula was obtained first in \cite{84QC}.

It should be noted that (\ref{eq:215f}) does not
assume that the composite system ${\bf S}+{\bf A}$ with the outcome
${\bf a}(t)\in\Delta$ is in the state
\begin{eqnarray}
\lefteqn{\rho_{{\bf S}+{\bf A}}(t+\Delta t|{\bf a}(t)\in\Delta)}
\quad\nonumber\\
&=&
\frac{(I\otimes E^{B}(\Delta))U(\rho(t)\otimes\sigma)U^{\dagger}
(I\otimes E^{B}(\Delta))}
{{\rm Tr}[(I\otimes E^{B}(\Delta))U(\rho(t)\otimes\sigma)U^{\dagger}]}
\label{eq:215g}
\end{eqnarray}
just after measurement.
In fact, such assumption is not correct,
since for any partition $\Delta=\Delta'\cup\Delta''$ the state 
$\rho_{{\bf S}+{\bf A}}(t+\Delta t|{\bf a}(t)\in\Delta)$ should be 
a mixture of
$\rho_{{\bf S}+{\bf A}}(t+\Delta t|{\bf a}(t)\in\Delta')$ and 
$\rho_{{\bf S}+{\bf A}}(t+\Delta t|{\bf a}(t)\in\Delta'')$ but 
it is not the case
for (\ref{eq:215g}).

In order to examine the mathematical properties of the selective state
change, for any Borel set $\Delta$ define the transformation $T_{\Delta}$
on the space $\tau c({\cal H})$ of trace class operators on ${\cal H}$ by 
\begin{equation}\label{eq:217a}
T_{\Delta}(\rho)
={\rm Tr}_{{\cal K}}[(I\otimes E^{B}(\Delta))U(\rho\otimes\sigma)U^{\dagger}]
\end{equation}
for any $\rho\in\tau c({\cal H})$ and we shall call 
the family $\{T_{\Delta}|\ \Delta\in{\cal B}({\bf R})\}$ the {\em operational 
distribution} of the apparatus ${\bf A}$, where ${\cal B}({\bf R})$ is the
collection of Borel sets in ${\bf R}$.  

Then, it is easy to check that the family 
$\{T_{\Delta}|\ \Delta\in{\cal B}({\bf R})\}$ 
satisfies the following conditions:
\begin{mathletters}

(i) For any Borel set $\Delta$, the map $\rho\mapsto T_{\Delta}(\rho)$ is
completely positive \cite{fn:225a}, i.e., we have
\begin{equation}\label{eq:217b}
\sum_{i,j=1}^{n}\langle\xi_{i}|T_{\Delta}(\rho_{i}^{\dagger}\rho_{j})
|\xi_{j}\rangle\ge0
\end{equation}
for any finite sequences
$\xi_{1},\ldots,\xi_{n}\in{\cal H}$  and 
$\rho_{1},\ldots,\rho_{n}\in\tau c({\cal H})$.

(ii) For any Borel set $\Delta$ and disjoint Borel sets $\Delta_{n}$ such that
$\Delta=\bigcup_{n}\Delta_{n}$, we have
\begin{equation}\label{eq:1230a}
T_{\Delta}(\rho)=\sum_{n}T_{\Delta_{n}}(\rho)
\end{equation}
for any $\rho\in\tau c({\cal H})$.

(iii) For any Borel set $\Delta$ and any $\rho\in\tau c({\cal H})$,
\begin{equation}\label{eq:1230c}
{\rm Tr}[T_{\Delta}(\rho)]={\rm Tr}[E^{A}(\Delta)\rho].
\end{equation}
\end{mathletters}

Mathematically, we shall call any map $\Delta\mapsto T_{\Delta}$ satisfying
conditions (i) and (ii) a {\em completely positive (CP) map valued measure}
and it is said to be {\em $A$-compatible} if condition (iii) holds.
We have shown that {\em the operational distribution of any model
apparatus $({\cal K},\sigma,U,B)$ measuring $A$ is an $A$-compatible CP 
map valued 
measure}.  The converse of this assertion was proved in \cite{84QC}
so that 
{\em for any $A$-compatible CP map valued measure $\Delta\mapsto T_{\Delta}$ 
there is a model apparatus $({\cal K},\sigma,U,B)$ measuring $A$ such that
its operational distribution is 
$\{T_{\Delta}|\ \Delta\in{\cal B}({\bf R})\}$}.

From (\ref{eq:215b}) and (\ref{eq:1230c}) we have 
\begin{equation}\label{eq:221b}
{\rm Pr}\{{\bf a}(t)\in\Delta\}={\rm Tr}[T_{\Delta}(\rho(t))],
\end{equation}
and from (\ref{eq:215f}) and (\ref{eq:217a}), if 
${\rm Pr}\{{\bf a}(t)\in\Delta\}>0$,
we have
\begin{equation}\label{eq:221c}
\rho(t+\Delta t|{\bf a}(t)\in\Delta)
=\frac{T_{\Delta}(\rho(t))}{{\rm Tr}[T_{\Delta}(\rho(t))]}.
\end{equation}
Thus, both the outcome distribution and the selective state change are 
determined by the operational distribution;
relations (\ref{eq:221b}) and (\ref{eq:221c}) were postulated for positive 
map valued measures first by Davies and Lewis \cite{DL70}.

For any real number $a$, let $\rho(t+\Delta t|{\bf a}(t)=a)$ be the
state at $t+\Delta t$ of ${\bf S}$ conditional upon ${\bf a}(t)=a$; 
if the object leads to the outcome ${\bf a}(t)=a$, it is
in the state $\rho(t+\Delta t|{\bf a}(t)=a)$ at $t+\Delta t$.
The {\em state reduction} caused by the apparatus ${\bf A}$ 
is the state change 
$\rho(t)\mapsto\rho(t+\Delta t|{\bf a}(t)=a)$ for all real number $a$.
The family $\{\rho(t+\Delta t|{\bf a}(t)=a)|\ a\in{\bf R}\}$ of states
is called the {\em posterior states} for the {\em prior state} $\rho(t)$.
According to the above definition, the state reduction and 
the selective state change are related by
\begin{eqnarray}
\lefteqn{{\rm Pr}\{{\bf a}(t)\in\Delta\}\rho(t+\Delta t|{\bf a}(t)\in\Delta)}
\quad\nonumber\\
&=&
\int_{\Delta}\rho(t+\Delta t|{\bf a}(t)=a){\rm Pr}\{{\bf a}(t)\in da\},
\label{eq:215c}
\end{eqnarray}
or equivalently 
\begin{equation}
T_{\Delta}(\rho(t))
=
\int_{\Delta}\rho(t+\Delta t|{\bf a}(t)=a){\rm Tr}[E^{A}(da)\rho(t)]
\label{eq:226a}
\end{equation}
for any Borel set $\Delta$.
It was shown in \cite{85CA} that the above formula determines 
the posterior states as a Borel function 
$a\mapsto\rho(t+\Delta t|{\bf a}(t)=a)$ 
from ${\bf R}$ to the space of density operators uniquely up to probability 
one.
To disregard irrelevant difference,
two families of posterior states are considered to be {\em identical} 
if they are identical on a set of outcomes with probability one.  
Then, the state reduction and the selective
state change are equivalent under relation (\ref{eq:215c}).
We conclude, therefore, that {\em two model apparatuses are statistically
equivalent if and only if they have the same operational distribution.}

Let $a\mapsto\rho_{a}$ be an arbitrary Borel function from ${\bf R}$
to the density operators.  
We shall consider the following problem:
Is there any model apparatus measuring a given observable $A$ such that
the measured object is left in the state $\rho_{a}$ just after
measurement with outcome $a$ regardless of the prior state $\rho(t)$. 
The affirmative answer to this problem is obtained as follows.
For any Borel set $\Delta$, let
$T_{\Delta}$ be the transformation of $\tau c({\cal H})$ defined by
\begin{equation}\label{eq:221a}
T_{\Delta}(\rho)=\int_{\Delta}\rho_{a}{\rm Tr}[E^{A}(da)\rho]
\end{equation}
for any $\rho\in\tau c({\cal H})$.
Then, it can be shown that the map $\Delta\mapsto T_{\Delta}$ is an 
$A$-compatible completely positive map valued measure \cite{85CC}.
It follows that there is a model apparatus $({\cal K},\sigma,U,B)$ measuring
$A$ such that its operational distribution is given by (\ref{eq:221a}).
By comparing (\ref{eq:226a}) and (\ref{eq:221a}), we have
$\rho(t+\Delta t|{\bf a}(t)=a)=\rho_{a}$ \cite{fn:303a}.
Thus, we have shown that {\em for any observable $A$ and 
any Borel function $a\mapsto\rho_{a}$
there is a model apparatus measuring $A$
such that the posterior states is
$\{\rho_{a}|\ a\in{\bf R}\}$ regardless of the prior state of the object.}

Now, we shall consider the following model of position measurement 
\cite{88MS}.
The object ${\bf S}$ is a one dimensional mass with position ${\hat x}$,
momentum ${\hat p}_{x}$, and Hamiltonian ${\hat H}_{{\bf S}}$.
The apparatus ${\bf A}$ measuring the object position ${\hat x}$
is another one dimensional mass with position ${\hat y}$,
momentum ${\hat p}_{y}$, and Hamiltonian ${\hat H}_{{\bf A}}$.
The Hilbert space of ${\bf S}$ is 
${\cal H}=L^{2}({\bf R}_{x})$, the $L^{2}$ space
of the $x$-coordinate, 
and the Hilbert space of ${\bf A}$ is ${\cal K}=L^{2}({\bf R}_{y})$.
The object-apparatus coupling is turned on from time $t$ to $t+\Delta t$.
Suppose that the time dependent total Hamiltonian 
${\hat H}_{{\bf S}+{\bf A}}(T)$
of  ${\bf S}+{\bf A}$ is taken to be
\begin{eqnarray}
{\hat H}_{{\bf S}+{\bf A}}(T)
&=&{\hat H}_{{\bf S}}\otimes I+I\otimes {\hat H}_{{\bf A}}\nonumber\\
& &\mbox{ }-K_{1}(T){\hat p}_{x}\otimes {\hat y}
+K_{2}(T){\hat x}\otimes{\hat p}_{y},
\label{eq:301a}
\end{eqnarray}
where the strengths of couplings, $K_{1}(T)$ and $K_{2}(T)$, satisfy
\begin{mathletters}
\begin{eqnarray}
K_{1}(T)&=&0\quad\mbox{if $T\not\in(t,t+\frac{\Delta t}{2})$},\\
K_{2}(T)&=&0\quad\mbox{if $T\not\in(t+\frac{\Delta t}{2},t+\Delta t)$},
\end{eqnarray}
\begin{equation}
\int_{t}^{t+\frac{\Delta t}{2}}K_{1}(T)dT=1,\quad
\int_{t+\frac{\Delta t}{2}}^{t+\Delta t}K_{2}(T)dT=1.
\end{equation}
\end{mathletters}
We assume that $\Delta t$ is so small that the system Hamiltonians 
${\hat H}_{{\bf S}}$ and
${\hat H}_{{\bf A}}$ can be neglected from $t$ to $t+\Delta t$.
By the Schr\"{o}dinger equation,
the time evolution of ${\bf S}+{\bf A}$ during the coupling is described 
by the unitary evolution operators
\begin{mathletters}
\begin{eqnarray}
U(t+\frac{\Delta t}{2},t)
&=&\exp \frac{i}{\hbar}({\hat p}_{x}\otimes {\hat y}),\\
U(t+\Delta t,t+\frac{\Delta t}{2})
&=&\exp -\frac{i}{\hbar}({\hat x}\otimes {\hat p}_{y}).
\end{eqnarray}
\end{mathletters}
Then, in the position basis we have
\begin{mathletters}
\begin{eqnarray}
\langle x,y|U(t+\frac{\Delta t}{2},t)|x',y'\rangle
&=&\langle x+y,y|x',y'\rangle,\\
\langle x,y|U(t+\Delta t,t+\frac{\Delta t}{2})|x',y'\rangle
&=&\langle x,y-x|x',y'\rangle,
\end{eqnarray}
\end{mathletters}
and hence
\begin{eqnarray}
&\langle x,y|U(t+\Delta t,t+\frac{\Delta t}{2})
U(t+\frac{\Delta t}{2},t)|x',y'\rangle& 
\nonumber\\
&\quad=\langle y,y-x|x',y'\rangle.&
\label{eq:226b}
\end{eqnarray}

Let $\xi$ be an arbitrary normalized state vector of the apparatus.
We shall show that the model ${\bf M}=({\cal K},\sigma,U,B)$ 
measures ${\hat x}$
with ${\cal K}=L^{2}({\bf R}_{y})$, $\sigma=|\xi\rangle\langle\xi|$, 
$U=U(t+\Delta t,t+\frac{\Delta t}{2})U(t+\frac{\Delta t}{2},t)$,
and $B={\hat y}$.
Suppose that ${\bf S}$ is in the state 
$\rho(t)=|\psi(t)\rangle\langle\psi(t)|$
at the time $t$, where $\psi(t)$ is a normalized state vector of ${\bf S}$.
Then, ${\bf S}+{\bf A}$ has the state vector $\Psi(t)=\psi(t)\otimes\xi$
at $t$ and the state vector $\Psi(t+\Delta t)=U\Psi(t)$ at $t+\Delta t$.
In this case, we have
\begin{equation}\label{eq:225b}
U(\rho(t)\otimes\sigma)U^{\dagger}
=|\Psi(t+\Delta t)\rangle\langle\Psi(t+\Delta t)|,
\end{equation}
and by (\ref{eq:226b}) we have
\begin{equation}\label{eq:225c}
\langle x,y|\Psi(t+\Delta t)\rangle
=\langle y|\psi(t)\rangle\langle y-x|\xi\rangle.
\end{equation}
From (\ref{eq:215a}) the outcome distribution is given by
\begin{eqnarray}
{\rm Pr}\{{\bf a}(t)\in\Delta\}
&=&
\langle\Psi(t+\Delta t)|I\otimes E^{{\hat y}}(\Delta)
|\Psi(t+\Delta t)\rangle\nonumber\\
&=&
\int_{\Delta}\,dy
\int_{{\bf R}}|\langle y|\psi(t)\rangle|^{2}|\langle y-x|\xi\rangle|^{2}\,dx
\nonumber\\
&=&
\int_{\Delta}|\langle y|\psi(t)\rangle|^{2}\,dy.
\end{eqnarray}
Thus, the outcome distribution coincides with the ${\hat x}$ distribution at
the time $t$, so that the model ${\bf M}$ measures ${\hat x}$.

Now, let $\phi$ be an arbitrary normalized state vector and suppose that
the apparatus preparation $\xi$ is such that 
$\langle y|\xi\rangle=\langle-y|\phi\rangle$.
Then, we shall prove that the state reduction caused by the
model ${\bf M}$ is given by
\begin{equation}\label{eq:225a}
\rho(t+\Delta t|{\bf a}(t)=a)
=e^{-ia{\hat p}_{y}/\hbar}|\phi\rangle\langle\phi|e^{ia{\hat p}_{y}/\hbar}
\end{equation}
for arbitrary prior state $\rho(t)$ of ${\bf S}$.

By (\ref{eq:217a}), (\ref{eq:225b}), and (\ref{eq:225c}),  we have
\begin{eqnarray*}
\lefteqn{\langle x|T_{\Delta}(|\psi(t)\rangle\langle\psi(t)|)|x'\rangle}\\
&=&
\int_{\Delta}
\langle x,a|\Psi(t+\Delta t)\rangle\langle\Psi(t+\Delta t)|x',a\rangle\,da\\
&=&
\int_{\Delta}\langle x-a|\phi\rangle\langle\phi|x'-a\rangle
|\langle\psi(t)|a\rangle|^{2}\,da\\
&=&
\left\langle x\left|
\int_{\Delta}e^{-ia{\hat p}_{y}/\hbar}
|\phi\rangle\langle\phi|e^{ia{\hat p}_{y}/\hbar}
\,\langle\psi(t)|E^{{\hat x}}(da)|\psi(t)\rangle\right|x'\right\rangle.
\end{eqnarray*}
Since $x$ and $x'$ are arbitrary, we have
\begin{eqnarray}
\lefteqn{T_{\Delta}(|\psi(t)\rangle\langle\psi(t)|)}\nonumber\\
&=&
\int_{\Delta}
e^{-ia{\hat p}_{y}/\hbar}|\phi\rangle\langle\phi|e^{iy{\hat p}_{y}/\hbar}
\,\langle\psi(t)|E^{{\hat x}}(da)|\psi(t)\rangle.
\end{eqnarray}
By the linearity of $T_{\Delta}$, we have
\begin{equation}
T_{\Delta}(\rho)
=\int_{\Delta}
e^{-ia{\hat p}_{y}/\hbar}|\phi\rangle\langle\phi|e^{ia{\hat p}_{y}/\hbar}
\,{\rm Tr}[E^{{\hat x}}(da)\rho]
\end{equation}
for any trace class operator $\rho$.
This shows that the model ${\bf M}$ has the operational distribution of
the form (\ref{eq:221a}) with 
$$
\rho_{a}
=e^{-ia{\hat p}_{y}/\hbar}|\phi\rangle\langle\phi|e^{ia{\hat p}_{y}/\hbar},
$$
and hence the state reduction caused by ${\bf M}$ is given by (\ref{eq:225a}).

In this paper, it is demonstrated that the measuring apparatus can 
be designed, in principle, so that the posterior states is controlled 
to be an arbitrary family of states.
This disproves the following interpretation of the uncertainty
principle: 
if the position is measured with noise $\Delta x$ then the
back action disturbs the momentum so that the momentum uncertainty
$\Delta p$ just after measurement is not less than the order of 
$\hbar/\Delta x$.

In \cite{98OSPPT}, it was proved that any measuring 
apparatus disturbs the probability distribution of every observable
not commuting with the measured observable in {\em some} prior state.
The present model suggests, however, that the assertion cannot be 
strengthened so that the apparatus disturbs it in {\em any} prior state,
as follows.
Suppose that the apparatus is prepared in a momentum 
eigenstate $|p\rangle$.
Then, the position measurement using {\bf M} leaves 
the object in the momentum eigenstate $|-p\rangle$ regardless of the outcome.
Thus, if the object is in the momentum eigenstate $|-p\rangle$ just
before measurement, the posterior state is the same state.
This shows that position can be measured without perturbing
the object momentum just before measurement.


\begin{thebibliography}{10}
\bibitem{fn:225a}
An alternative but equivalent condition is
that there is a sequence $\{A_{n}\}$ of bounded operators on ${\cal H}$
such that $T_{\Delta}(\rho)=\sum_{n}A_{n}\rho A_{n}^{\dagger}$ 
for all $\rho\in\tau c({\cal H})$.
\end{thebibliography}
\end{document}